\documentclass[notitlepage]{article}
\usepackage{amssymb}

\newtheorem{example}{Example}[section]
\newtheorem{remark}{Remark}[section]
\newtheorem{theorem}{Theorem}[section]
\newtheorem{corollary}{Corollary}[section]

\begin{document}

\pagestyle{plain}

\title{Geometry and integrability of Euler--Poincar\' e--Suslov equations}

\author{Bo\v zidar Jovanovi\' c \\
\\
Mathematisches Institut, LMU \\
Theresienstr. 39, D-80333 M\" unchen, Germany\\
e-mail:  jovanov$@$rz.mathematik.uni-muenchen.de\\
\\
Matemati\v cki Institut, SANU\footnote{permanent address}\\
Kneza Mihaila 35, 11000 Beograd, Serbia, Yugoslavia \\
e-mail: bozaj$@$mi.sanu.ac.yu 
}
\date{}
\maketitle

\begin{abstract}
We consider nonholonomic geodesic flows 
of  left-invariant metrics and left-invariant nonintegrable
distributions on  compact connected Lie groups.
The equations of geodesic flows are reduced to the 
Euler--Poincar\' e--Suslov equations on the corresponding Lie algebras.
The Poisson and symplectic structures give raise to 
various algebraic constructions of the integrable Hamiltonian systems.
On the other hand, nonholonomic systems are not Hamiltonian and
the integration methods for nonholonomic systems are much less
developed. In this paper, using chains of subalgebras,
we give constructions that lead
to a large set of first integrals and to integrable
cases of the Euler--Poincar\' e--Suslov equations.
Further, we give examples of nonholonomic geodesic flows
that can be seen as a restrictions of integrable sub-Riemannian
geodesic flows.
\end{abstract}

\newpage

\section{Introduction}

In this paper we are interested in the geometry and integrability of 
nonholonomic geodesic flows 
of  left-invariant metrics and left-invariant nonintegrable
distributions on  compact Lie groups.
The equations of geodesic flows are reduced to the 
Euler--Poincar\' e--Suslov equations on the corresponding Lie algebras.
These systems are natural generalizations to Lie algebras
of the Suslov nonholonomic rigid body problem
and were introduced independently by Kozlov \cite{Koz} and 
Koiller \cite{Koi}.
The known integrable cases are given by
Fedorov and Kozlov \cite{FK} and author \cite{Jo}.

In the recent years appears many papers concerning
geometrical formulation of the nonholonomic mechanics.
For instance, see \cite{Koi, BKMM, Ma} and references therein.
However,
since nonholonomic systems do not admit a Poisson structure,
the integration theory of the constrained mechanical systems
is much less developed then for the unconstrained.
We mention Chaplygin's results \cite{Ch}.
By the use of an invariant measure, he
had given some of the most interesting examples of the
solvable nonholonomic systems
and noticed that  the phase space could be foliated on invariant tori,
placing these  systems
together with integrable Hamiltonian systems.
The methods of integration of systems with an invariant measure, 
as well as  illustrative integrable examples can be found in   
\cite{AKN, VV, Ze, He}.

Now, we shall briefly describe  the results and outline of this paper.

In sections 1 and 2 we shall give basic definitions and notation.

One of the well known ways for studying
the integrability of Riemannian geodesic flows is by using
certain filtrations of Lie algebras (see \cite{Th, Mik, PS, Ba, BJ}).
By taking appropriate chains of subalgebras, we can get
integrable cases of  the 
Euler--Poincar\' e--Suslov equations.

In section 3 we shall consider the case when the  
left-invariant nonintegrable distribution is an invariant subspace of 
left-invariant metric.
Then reduced system has an invariant measure.
We shall construct integrable examples 
by using chains of the form $K\subset H\subset G$, where $(G,H)$ is a symmetric
pair. The obtained systems are generalizations of the Fedorov--Kozlov
integrable case.

In section 4 we deal with an arbitrary chain 
$G_0\subset G_1 \subset \dots \subset G_n=G$.
This gives us opportunity to construct exactly solvable examples 
without an invariant measure as well.

In some cases the  interesting phenomena arises:
the nonholonomic geodesic flow could be seen as a 
restriction of Hamiltonian flow on the whole, unconstrained
phase space.
We shall present here two families of integrable sub-Riemannian
geodesic flows which restrict to the our non-Hamiltonian problem
 (section 6).

\section{Nonholonomic geodesic flows}

Let $(Q,(\cdot,\cdot))$ be $n$--dimensional Riemannian manifold with 
Levi--Civita connection $\nabla$. Let $\mathcal D$ be the nonintegrable 
$\rho$--dimensional distribution distribution of the tangent bundle. 
The distribution can be defined by 
$n-\rho$ independent one forms $\alpha_i$ in the following way:
$$
\mathcal D_q=\{ \xi\in T_qQ, \; \alpha_i(\xi) =0,\;
i=1,\dots,\rho \}.
$$

The smooth path $\gamma(t), t\in\Delta$ is called {\it admissible} 
(or allowed by constraints) if 
velocity $\dot\gamma(t)$ belongs to ${\mathcal D}_{\gamma(t)}$ for all $t\in\Delta$.
There are two approaches to
define the geodesic lines among admissible paths: by 
induced connection as a "straightest" lines 
and by variational principle as a "shortest" lines.
We shall deal with the first approach which arises from
mechanics. The admissible path $\gamma(t)$ is called 
{\it a nonholonomic geodesic line} if it
satisfied {\it d'Alambert--Lagrange equations}:
\begin{equation}
\pi(\nabla_{\dot\gamma(t)}\dot \gamma(t))=0,
\label{0.1}
\end{equation}
where $\pi: T_q Q\to {\mathcal D}_q$, $q\in Q$ is the orthogonal projection.

Let $\{\cdot,\cdot\}$ be canonical Poisson brackets on $T^*Q$.
By $X_f$ we shall denote the {\it Hamiltonian vector field} of the function
$f$ ($dg(X_f)=\{g,f\}$, for all $g:T^*Q\to\Bbb{R}$).
Let $\mathcal M$ be the {\it constraint submanifold} in the phase space $T^*Q$:
$$
\mathcal M=\{(p,q)\in T^*Q, \; p\in g_q(\mathcal D_q)\subset T^*_q Q\},
$$
where we consider scalar product $(\cdot,\cdot)_q$ as the 
mapping $g_q: T_q Q \to T_q^*Q$.
Taking for the Hamiltonian $h(p,q)=\frac12 p(g^{-1}_q p)$, $p\in T^*_q Q$,
we can write (\ref{0.1}) in the following form:
\begin{equation}
\dot x=X_h(x)+\sum_{i=1}^\rho \lambda_i \mathrm{vert} \alpha_i\vert_{\sigma(x)},
\label{d1.1}
\end{equation}
where $\mathrm{vert}\alpha_i\vert_{\sigma(x)}\in T_x(T^*Q)$ is 
"vertical" lift of $\alpha_i\vert_{\sigma(x)}\in T^*_{\sigma(x)}Q$,
$\sigma: T^*Q\to Q$ is the natural projection
and Lagrange multipliers $\lambda_i$ are chosen such that the phase trajectory $x(t)$ belongs to 
$\mathcal M$. 
In canonical coordinates $x=(p,q)$  equations (\ref{d1.1})
are:
\begin{equation}
\dot p=-\frac{\partial h(p,q)}{\partial q}+ 
\sum_{i=1}^\rho \lambda_i \alpha_i(q),
\quad
\dot q=\frac{\partial h(p,q)}{\partial p}.
\label{d1.2}
\end{equation}

It is important to note that the equations
are not Hamiltonian (they are Hamiltonian with respect to the
almost-Poisson brackets on $\mathcal M$ \cite{Ma, CLM}) and that 
the Hamiltonian function is conserved.

Sufficient condition for the integration
of the non-Hamiltonian system
\begin{equation}
\dot x=f(x), \quad x\in\mathbb{R}^m,
\label{d1.3}
\end{equation}
is existence of $m-1$ independent
integrals, or one integral less
in the case of the existence of an invariant measure.

Suppose that the system of equations (\ref{d1.3}) has an invariant
measure and $m-2$ first integrals $F_1,\dots,F_m$.
If $F_1,\dots,F_m$ are independent on the invariant set
$M_c=\{x\in\mathbb{R}^m,\; F_i(x)=c_i, \; i=1,\dots,m-2\}$ then
the solution of  (\ref{d1.3}) lying on $M_c$ can be found by
quadratures (the Jacobi theorem).
Moreover,
if $L_c$ is a compact connected component of $M_c$
and $f(x) \ne 0$ on $L_c$ then $L_c$ is diffeomorphic to a two-torus;
one can find angular coordinates $\varphi_1,\varphi_2$ on $L_c$ in which
equations  (\ref{d1.3}) take the form
similar as in the Liouville theorem:
$$
\dot\varphi_1=\frac{\omega_1}{\Phi(\varphi_1,\varphi_2)},\quad
\dot\varphi_2=\frac{\omega_2}{\Phi(\varphi_1,\varphi_2)},
$$
where $\omega_1,\omega_2$ are constant and $\Phi$ is a smooth
positive $2\pi$--periodic function in $\varphi_1,\varphi_2$
(see \cite{AKN}).

Therefore  
it is natural to call a non-Hamiltonian system {\it integrable}
if it can be integrated by the above procedure; 
or more generally
(as it was point out in \cite{VV}),
if the trajectories of the system belong to invariant tori with
dynamic of the form
\begin{equation}
\dot \varphi_1=\frac{\omega_1}{\Phi(\varphi_1,\dots,\varphi_k)},\dots,
\dot \varphi_k=\frac{\omega_k}{\Phi(\varphi_1,\dots,\varphi_k)}.
\label{d1.4}
\end{equation}

The flow of (\ref{d1.4}) is unevenly winding and
admit the invariant measure
$\mu(D)=\int_D \Phi d\varphi_1\wedge\dots\wedge d\varphi_k$.
Note that it is shown in \cite{AKN} that
for almost all frequencies
$\omega_1,\dots,\omega_k$, by smooth change of variables
$$
\mathbb{T}^k\{\varphi_1,\dots,\varphi_k\}\to
\mathbb{T}^k\{\theta_1,\dots,\theta_k\},
$$
equations (\ref{d1.4}) can be reduced to the form 
$\dot \theta_i=\Omega_i=\omega_i/\Pi$,
$i=1,\dots,k$, 
where $\Pi$ denotes the total measure of $\mathbb{T}^k$.

\section{ Euler--Poincar\' e--Suslov equations}

Now, let $Q$ be a compact connected Lie group  $\mathfrak G$ with
Lie algebra $G=T_e\mathfrak G$.
In what follows we shall identify
$G$ and $G^*$ by $Ad_\mathfrak G$ invariant
scalar product $\langle \cdot,\cdot\rangle$;
$T\mathfrak G$ and $T^*\mathfrak G$ by bi-invariant metric on $\mathfrak G$.

We shall consider left-invariant distributions. Let
$$
D=\{\omega\in G,\; \langle \omega,a^i \rangle =0, \; i=1,\dots,\rho \}\subset G 
$$
be the restriction of the left-invariant distribution $\mathcal D$ 
to the Lie algebra $G$, for some 
constant, linearly independent  vectors $a^i$ in $G$.
The distribution is nonintegrable if and only if $D$ is not a subalgebra.
From the invariance, we have that $\mathcal D_g=g\cdot D$.

Let $I: G\to G$ be a symmetric, positive definite (with respect to $\langle \cdot,\cdot\rangle$)
operator that induces left-invariant metric:
$$
(\eta_1,\eta_2)_g=\langle I(\omega_1),\omega_2\rangle,
\; \eta_1,\eta_2\in T_g\mathfrak G, \; 
\omega_1,\omega_2\in G, \; \eta_1=g\cdot\omega_1,\; \eta_2=g\cdot \omega_2.
$$
Let $M$ be the restriction of the constraint 
submanifold $\mathcal M$ to $G$.
Then $M=I(D)$ and $\mathcal M_g=g\cdot M$.
The Hamiltonian of the geodesic flow
is  function
$h: T\mathfrak G\to \Bbb{R}$ obtained from 
{\it reduced Hamiltonian function}
$H(x)=\frac12 \langle A(x),x\rangle$ by left-translations
(here  $A=I^{-1}$).

In a such notation,
the equations (\ref{d1.1}) are reduced to:
\begin{equation}
\dot x = [x,\nabla H(x)]+ \sum_{i=1}^{\rho} \lambda_i a^i=[x, A(x)] + \sum_{i=1}^{\rho} \lambda_i a^i, 
\label{0.3}
\end{equation}
where Lagrange multipliers are chosen such that 
$x$ belongs to $M=I(D)$, i.e.
such that $\omega=A(x)$ belongs to $D$:
$$
\langle A(x),a^i \rangle =0, \quad i=1,\dots, \rho.
$$

According to Fedorov and Kozlov \cite{FK} we shall call these equations the 
{\it Euler--Poincar\' e--Suslov equations},
as a generalization of the Suslov nonholonomic rigid body problem.
They have a quite different nature of the corresponding Euler--Poincar\' e equations $\dot x=[x,A(x)]$. 
For instance, in general,
they do not have a smooth invariant measure (see \cite{Koz}).
The  equations  (\ref{0.3}) could be seen also as a reduced
equations from a point of view of a reduction of nonholonomic
systems with symmetries given in \cite{BKMM, Ma}.

Nonholonomic geodesic lines $g(t)$ are solution of the kinematic
equation $g^{-1}(t)\cdot \dot g(t)=\omega(t)=A(x(t))$, 
where $x(t)$ are solutions of (\ref{0.3}).
In other words, the following diagram commutes:
\begin{equation}
\begin{array}{ccccc}
     &  & {\mathcal P}^t &  &   \\
&{\mathcal M} &  \rightarrow     & {\mathcal M}& \\
 \Lambda & \downarrow & &  \downarrow & \Lambda\\
& M & \rightarrow & M & \\
& & P^t & & \\
\end{array}
\label{0.4}
\end{equation}
Here,
${\mathcal P}^t$  and $P^t$ are phase flows of the
nonholonomic geodesic flow and Euler--Poincar\' e--Suslov equations;
$\Lambda$ maps $p=g\cdot x \in T_g\mathfrak G$ to $x\in G$.
If Euler--Poincar\' e--Suslov equations are integrable
we shall say that nonholonomic geodesic flow
is integrable in the sense of the factorization (\ref{0.4}).

The reduced Hamiltonian function $H(x)=\frac12\langle x,A(x) \rangle$
is the first integral of system (\ref{0.3}).
This follows from the conservation of energy in the natural 
mechanical nonholonomic systems with linear constraints.
One may also prove this fact by a direct computation of the
time derivative of $H(x)$ along the vector field defined by (\ref{0.3}).
Namely, integral $F$ of Euler--Poincar\' e equations $\dot x=[x,A(x)]$ is the integral
of Euler--Poincar\' e--Suslov equations (\ref{0.3}) if and only if:
\begin{equation}
\sum_i \lambda_i \langle \nabla F(x),a^i\rangle\vert_{x\in M}=0.
\label{d2.1}
\end{equation}

\section{Symmetric pairs}

Let $L$ be the subspace of $G$ spanned by
$a^i,\; i=1,\dots,\rho$.
In this section we shall consider the case when $A$ preserve
the orthogonal decomposition $G=L+D$, i.e., $A=A_L+A_D$, where 
$A_L:L\to L$, $A_D:D\to D$ are positive definite operators. 
Then $M=I(D)=D$ and
we can write (\ref{0.3}) in the following way:
\begin{equation}
\dot \xi=[\xi,A_D(\xi)]_D, \quad \xi\in D
\label{1.1}
\end{equation}
(by $x_K$ we denote the orthogonal projection of $x$ to the linear space $K$).

The equations (\ref{1.1}) preserve the standard measure on $D$.
Also the {\it constrained reduced Hamiltonian function} 
$H_D=\frac12\langle \xi,A_D(\xi)\rangle$
and invariant 
$F(\xi)=\langle \xi,\xi \rangle$ are always
first integrals of the system.
Note that by (\ref{d2.1}), in general, 
the invariant $F(x)=\langle x,x \rangle$ is not the integral
of (\ref{0.3}).

\begin{example}{\rm We have, at least, following integrable cases:

\item{\bf 1.} If $A_D=s\cdot Id_D$, $s\in \mathbb{R}$
 then the solution of (\ref{1.1}) are $\xi=const$. In this
case the constraints have no influence to the motion (the Lagrange
multipliers in (\ref{0.3}) are equal to zero). The nonholonomic geodesic lines 
are simply given with
$g(t)=g_0\exp(\eta t)$, $\eta\in D$, $g_0\in \mathfrak G$. 
At the same time these lines are geodesic lines
of the left-invariant metric induced by $A=A_L+s\cdot Id_D$. 
From now on we shall suppose that $A_D \neq s\cdot Id_D$.

\item{\bf 2.} $\dim D=2$. Then the solution of (\ref{1.1}) 
are $\xi=const$. 

\item{\bf 3.} $\dim D=3,4$. Then the solutions of (\ref{1.1}) belong to 
the intersections of the spheres and ellipsoids: 
$$
M_{c_1,c_2}=\{\xi\in D, \; H_D(\xi)=\frac12\langle \xi,A_D(\xi)\rangle=c_1,
\; F(\xi)=\langle \xi,\xi\rangle=c_2\}.
$$
For $\dim D=3$ general trajectories are periodic.
If $\dim D=4$ then (\ref{1.1}) could be integrated by Jacobi 
theorem as a system with an invariant measure
(note that general connected
components of invariant submanifolds $M_{c_1,c_2}$
are 2--dimensional spheres).

\item{\bf 4.} 
Recall that $(G,H)$ is called a 
{\it symmetric pair} if the following condition is satisfied:
$$
[H,H]\subset H,\quad [H,V]\subset V,\quad [V,V]\subset H,
$$
where $V$ is the orthogonal complement of $H$.
If there is a subalgebra $H\subset G$ such that $(G,H)$ is
a symmetric pair and that $H\subset L \subset G$, then $[D,D]\subset H$ and
$[D,D]_D=0$. Therefore the solution of (\ref{1.1}) are $\xi=const$.

}
\end{example}

In the cases 2 and 4 above, nonholonomic geodesic lines
are $g(t)=g_0\exp(\eta t)$, $\eta\in D$, $g_0\in\mathfrak G$,
but these lines  not need to be geodesic lines of the 
left-invariant metric induced by $A$. 

Motivated with the last example,
let us consider the chain of subalgebras:
$$
K\subset H\subset G,
$$
such that $(G,H)$ is a symmetric pair and that $H$ is not a subspace of $L$.
Let $G=H+V$ be the orthogonal decomposition. We can
consider the adjoint representation of $K$ on the 
linear space $V$: $\eta\in K \mapsto 
[\eta,\cdot]\in End(V)$. Decompose $V$ on irreducible subspaces $V=V_0+V_1+\dots+V_m$.
Here $V_0$ denotes the subspace with trivial representation.

Let $D$ be of the form:
$$
D=U+W_0+W_1+\dots+W_m, \quad U=D\cap K,\quad W_k=D\cap V_k.
$$

Since $H\nsubseteq L$ we have that $\dim U \ge 1$.
Let $\dim W_k\ge 1$, $k=1,\dots,n$, $\dim W_k=0$, $k>n$.

\begin{theorem}
Suppose that operator $A_D$ preserve the decomposition $D=U+W_0+W_1+\dots+W_n$ and that
$A_D\vert_U=s\cdot Id_{U}$, $s\in \mathbb{R}$. 
Then equations (\ref{1.1}), besides functions
$H_D(\xi)=\frac12\langle A_D(\xi),\xi\rangle$ and
$F(\xi)=\langle \xi,\xi \rangle$,
have a set of the first integrals, the projection of $\xi$ to $W_0$: $F_0(\xi)=\xi_{W_0}$ 
and functions:
$$
F_k(\xi)=\langle B_{W_k} (\xi_{W_k}),\xi_{W_k}\rangle, \quad k=1,\dots,n
$$ 
where $B_{W_k}=A_D\vert_{W_k}-s\cdot Id_{W_k}$. 
\end{theorem}

{\it Proof.}
Let $A_W=A_D\vert_W$, $A_{W_k}=A_D\vert_{W_k}$, $W=W_0+\dots+W_n$.
The equations (\ref{1.1}) have the form:
\begin{equation}
\frac{d}{dt}(\xi_U+\xi_W)=[\xi_U+\xi_W, s\xi_U+A_W(\xi_W)]_D.
\label{1.5}
\end{equation}
From $[U,V]\subset V$, $[W,W]\subset H$ and
(\ref{1.5}) we get:
$$
\dot \xi_U=[\xi_W,A_W(\xi_W)]_U, 
$$
$$
\dot \xi_W=[\xi_U,A_W(\xi_W)-s\xi_W]_W.
$$

Since $A_W$ preserve the decomposition $W=W_0+W_1+\dots+W_n$, the second equation is 
separated on $n+1$ equations:
\begin{equation}
\dot \xi_{W_0}=0,\quad \dot \xi_{W_k}=[\xi_U,B_{W_k}(\xi_{W_k})]_{W_k}, 
\quad k=1,\dots,n.
\label{1.7}
\end{equation}
It is clear that $F_k$ are integrals of (\ref{1.7}).
Note that the invariant $F=\langle \xi,\xi \rangle$ is dependent of 
functions $H_D$ and $F_k$, $k=0,\dots,n$. 

\begin{corollary}
If operators $B_{W_k}$ are positive definite and $c_{n+1}$ 
satisfied inequality:
$$
c_{n+1}> \vert c_0 \vert^2 + \sum_{k=1}^n \frac{c_k}{b_k}, \quad
b_k=\min_{\vert \xi_{W_k} \vert=1} \langle B_{W_k}(\xi_{W_k}),\xi_{W_k}\rangle,
$$
then invariant subspaces:
$$
M_{c}=\{ \xi\in D,\; \xi_{W_0}=c_0, F_1(\xi)=c_1,\dots, F_n(\xi)=c_n, F(\xi)=c_{n+1}\}
$$
are diffeomorphic  to the product of spheres:
$$
S^{\dim W_1-1}\times \dots\times S^{\dim W_n-1} \times S^{\dim U-1}.
$$
In particular, if $\dim W_k \le 2$, $k=1,\dots,n$, $\dim U =1$, 
then $M_{c}$ is diffeomorphic
to the disjoint union of $g\le n$ dimensional tori
with quasi-periodic dynamic (\ref{d1.4}).
\end{corollary}

{\it Proof.}
The first part of the corollary follows from the relations:
$$
\vert \xi_U \vert^2=c_{n+1}-\vert c_0\vert^2-\sum_{k=1}^n\vert \xi_{W_k}\vert^2,
\quad \vert \xi_{W_k} \vert^2 \le \frac{c_k}{b_k}, \quad 
\langle B_{W_k}(\xi_{W_K}),\xi_{W_k} \rangle =c_k.
$$

Let $\dim W_k=2$, $k=1,\dots,g$, $\dim W_k=1$, $k=g+1,\dots,n$.
Let $\phi_k \;\mathrm{mod}\; 2\pi$ be the angular variables of ellipses 
$F_k=c_k$, $k=1,\dots,g$.
Let $\xi_U=\Phi\cdot \eta$, $\Phi\in \mathbb{R}$, $\eta\in U$, 
$\vert\eta\vert=1$.
Then $\Phi$ can be expressed in terms of the $\phi_k$ up to the sign:
$$
\Phi=\Phi(\phi_1,\dots,\phi_g)=
\pm\sqrt{c_{n+1}-\vert c_0 \vert^2-
\sum_{k=1}^g\vert \xi_{W_k}(\phi_k)\vert^2
-\sum_{k=g+1}^n \frac{c_k}{b_k}}\neq 0.
$$

The sign of $\Phi$ is  determined by 
the choice of the connected  component $\mathbb{T}^g$ of
$M_{c}$.

By substitution of 
$\xi_{W_k}=\xi_{W_k}(\phi_k)$, $k=1,\dots,g$, 
$\xi_U=\Phi(\phi_1,\dots,\phi_g)\cdot\eta$ to  (\ref{1.7}),
we obtain that angular
variables satisfied the following type equations:
\begin{equation}
\dot \phi_1=\Phi(\phi_1,\dots,\phi_g)f_1(\phi_1),\dots,
\dot \phi_g=\Phi(\phi_1,\dots,\phi_g)f_g(\phi_g).
\label{1.11}
\end{equation}
The system (\ref{1.11}) can be integrated in term of the new time $\tau$
defined by: $d\tau=\Phi(\phi_1,\dots,\phi_g)dt$.

If all $f_k$ are different from zero, then we can introduce 
angular variables $\varphi_k$ by averaging:
$$
\varphi_k=\varphi_k(\phi_k)={\omega_k} 
\int_0^{\phi_k}\frac{ds}{f_k(s)},\quad
\omega_k={2\pi}\left[{\int_0^{2\pi}\frac{ds}{f_k(s)}}\right]^{-1},
\quad k=1,\dots,g,
$$
in which (\ref{1.11}) takes the required form
\begin{equation}
\dot \varphi_1=\frac{\omega_1}{\Phi^{-1}},\dots,
\dot \varphi_g=\frac{\omega_g}{\Phi^{-1}}.
\label{1.13}
\end{equation}

\begin{remark}
{\rm
The frequencies $\omega_i$ depend only of the metric $A$.
If the trajectories are periodic on one torus,
they are periodic on the rest of the tori as well. 
}
\end{remark}

\begin{remark} {\rm
We have conditions  $\dim W_i=\dim V_i=2$, $\dim U=\dim K=1$ taking:
$$
G=so(n) = \left(
\begin{array}{cc}   U & W \\
                    -W^t & L \\
               \end{array}
\right), \; U=so(2), \; L=so(n-2), \;D=U+W.
$$
So, we can see the corollary as a generalization of 
the Fedorov--Kozlov integrable case \cite{FK}. }
\end{remark}

\begin{remark}
{\rm
Besides of preserving the standard measure in $D$,
we can easily rewrite system (\ref{1.1}) in the "Hamiltonian form":
\begin{equation}
\dot F=\{F,H_D\}_D, \quad F:D\to\mathbb{R},
\label{1.14}
\end{equation}
where $H_D=\frac12\langle\xi,A_D(\xi)\rangle$ and $\{\cdot,\cdot\}_D$
are almost-Poisson brackets defined by:
\begin{equation}
\{F_1,F_2\}_D(\xi)=\langle \xi, [\nabla F_2(\xi),\nabla F_1(\xi)] \rangle, 
\quad F_1,F_2: D \to \mathbb{R}.
\label{1.15}
\end{equation}
These brackets are bi-linear, skew-symmetric and satisfy the Leibniz rule.
In the general case, they do not satisfy the Jacobi identity.
For $D=G$ these are the usual Lie-Poisson brackets on the Lie algebra $G$.

As for the Poisson brackets, we can define central (or Casimir) functions of the
brackets (\ref{1.15}).
These are the functions $F$ that commute with all functions 
$F:D\to \mathbb{R}$.
Obviously,  they are integrals of the system (\ref{1.1}).
The example of the  central function is $F(\xi)=\langle \xi,\xi \rangle$. 

More about the  almost-Poisson setting  for the 
nonholonomic systems can be found in \cite{Ma, CLM}.
}
\end{remark}

\section{Chains of subalgebras}

Suppose we are given a chain of connected compact subgroups:
$$
\mathfrak G_0\subset \mathfrak G_1 \subset \dots \subset \mathfrak G_n=
\mathfrak G
$$
and the corresponding chain of subalgebras:
$$
G_0 \subset G_1 \subset \dots \subset G_n=G.
$$
Let $G_{i}=G_{i-1}+V_i$ be the orthogonal decompositions.
Then 
$$
G_i=G_0+V_1+\dots+V_i.
$$ 
Following \cite{Bo, Mik}, consider $A$ of the form:
\begin{equation}
A=A_0+s_1\cdot Id_{V_1}+\dots+s_n\cdot Id_{V_n}, \quad s_i>0,\; i=1,\dots,n,
\label{2.2}
\end{equation}
where $A_0$ is a symmetric positive operator 
defined in the subalgebra $G_0$. 

Suppose that $D$ has orthogonal decomposition:
\begin{equation}
D=D_0+D_1+\dots+D_n,
\label{2.3}
\end{equation}
$$
D_k=\{\omega_k\in V_k,\; \langle a_k^i,\omega_k \rangle =0, \; i=1,\dots,\rho_k\}.
$$
Then $D_k$, $k>0$ are invariant subspaces of $A$.
By $x_{k}$ denote the orthogonal projection of $x$ to $D_k$, $k>0$;
and by $x_0$ denote the orthogonal projection to $G_0$.

\begin{theorem}
The  Euler--Poincar\' e--Suslov equations (\ref{0.3}), 
with $D$ and operator $A(x)$ of the form (\ref{2.2}) and (\ref{2.3}),
are equivalent to the  Euler--Poincar\' e--Suslov equations on the 
Lie subalgebra $G_0$:
\begin{equation}
\dot x_0 = [x_0, A_0(x_0)] + \sum_{i=1}^{\rho_0} \mu_i a^i_0, 
\label{2.5}
\end{equation}
$$
\langle A_0(x_0),a^i_0 \rangle =0, \quad i=1,\dots, \rho_0,\\
$$
together with a chain of linear differential equations on the subspaces $D_k$:
\begin{equation}
\dot x_{k}=[x_{k},
A_0(x_0)-s_kx_0+(s_1-s_k)x_{1}+\dots+(s_{k-1}-s_k)x_{k-1}]_{k}.
\label{2.6}
\end{equation}
\end{theorem}

{\it Proof.}
Let $x_{V_k}$ be the orthogonal projection of $x$ to $V_k$.
The simple computations show:
\begin{equation}
\begin{array}{c}
[x,A(x)]=[x_{0}+x_{V_1}+\dots+x_{V_n},A_0(x_0)+s_1x_{V_1}+\dots+s_nx_{V_n}]\\
=[x_{0},A_0(x_{0})]+[x_{V_1},A_0(x_{0})-s_1x_{0}]+\dots+\\
+[x_{V_n},A_0(x_{0})-s_nx_{0}+(s_1-s_n)x_{V_1}+\dots+
(s_{n-1}-s_n)x_{V_{n-1}}].
\label{2.7}
\end{array}
\end{equation}
Taking into account (\ref{2.3}), (\ref{2.7}) and relations
$[G_{k-1},V_k]\subset V_k$, $k=1,\dots,n$,
after orthogonal projection of (\ref{0.3}) to the 
linear space $G_0+D_1+\dots+D_n$  the theorem follows.

\begin{remark} {\rm
Let $D_k=V_k$ and
let $\mathcal{F}_k$ be the algebra of polynomials 
 that are constant on the
orbits of the adjoint action of the group $\mathfrak G_{k-1}$ on  $V_k$. 
It can be easily seen that they are integrals of (\ref{2.6}).
The number of functionally independent polynomials in 
$\mathcal{F}_k$ is equal to:
$$
\dim V_k- \dim G_{k-1} + 
\min_{x_k\in V_k}\dim\{y\in G_{k-1}, \, [y,x_{k-1}]=0\}.
$$
}\end{remark}

If Euler--Poincar\' e--Suslov equations (\ref{2.5}) on $G_0$ are solvable
then the integration of original equations (\ref{0.3}) reduced to successive
integration of the chain of linear dynamical systems (\ref{2.6}) for $k>0$.

The most simplest case is when the solutions of (\ref{2.5}) are $x_0=const$.
Then the vector $x_1$ satisfied a linear equation with
constant coefficient and it is a elementary functions of the time $t$.
This is happened if $A_0=Id_{G_0}$ or if $G_0$ is a commutative subalgebra
(see also example 3.1)
In particular if $\dim D_0=0$ then we have that $x_0=0$.
In that case $\dot x_1=0$ and $x_2$ is a 
elementary functions of the time $t$.
This case will be treated again in the last section (example 6.1).

\begin{example}{\rm In addition, we can obtain exactly solvable cases
without an invariant measure.
For instance, take the chain: 
$$
so(3)\subset so(4) \dots \subset so(n).
$$
Let $D_0=\{\omega_0 \in so(3), \; \langle\omega_0,a_0 \rangle=0\}$.
If $a_0$ is not an  eigenvector of the $A_0$ then 
equations:
\begin{equation}
\dot x_0=[x_0,A_0(x_0)]+\lambda a_0,\quad\langle A(x_0),a_0\rangle=0,
\label{2.9}
\end{equation}
have no invariant measure \cite{Koz}.
After identification 
$(so(3),[\cdot,\cdot])=(\mathbb{R}^3,\times)$, 
the system (\ref{2.9}) 
describe the rotation of a rigid body
fixed at a point and subject to the constraint: the
angular velocity $\vec\omega_0$ is orthogonal to
the fixed vector in body coordinates $\vec a_0$.
This nonholonomic problem was solved by Suslov \cite{Su}.

The phase space $M_0=I_0(D_0)$ has the following property:
there is a asymptotic line $l$ such that 
$\lim_{t\to\pm\infty} P_0^t(\Omega_0)\subset l$, where $P^t_0$ is a phase
flow on $M_0$ and $\Omega_0$ is any subset of $M_0$.
Then the phase flow $P^t$ of (\ref{2.5}), (\ref{2.6}) has
asymptotic hyper-plane and also has no invariant measure.

However, for $\dim D_k\le 2$ we can easily integrated corresponding systems (\ref{2.6}).
Suppose that we know $y_{k-1}(t)=A_0(x_0(t))-s_kx_0(t)+(s_1-s_k)x_{1}(t)+\dots+
(s_{k-1}-s_k)x_{k-1}(t)$ as a function of time.
If $\dim D_k=1$ then $x_k=const$. For $\dim D_k=2$, 
let $\phi_k$ be the angular variable of the circle 
$I_k=\langle x_k,x_k \rangle=c_k$. Then it can easily be checked 
that equations 
$\dot x_k=[x_k,y_{k-1}(t)]_k$ get the form
$
\dot \phi_k=f_k(t),
$
where $f_k(t)$ is some known function of time.
Note that connected components of invariant submanifolds 
$H=h,I_1=c_1,\dots,I_{n-3}=c_{n-3}$
are diffeomorphic to tori, but the dynamic on tori 
is quite different from (\ref{1.13}).}
\end{example}

\section{Reconstruction}

The integrability of
Euler--Poincar\' e equations $\dot x=[x,A(x)]$ implies 
{\it non-commu\-tative integrability} of unconstrained geodesic flow ---
the phase space $T\mathfrak G$ is foliated on 
$d \le \dim \mathfrak G$ 
dimensional invariant isotropic tori 
with quasi-periodic dynamics
(see \cite{MF}).
In the nonholonomic case there is no Poisson structure.
In order to precisely describe dynamic on the whole phase space $\mathcal M$ 
we have to solve kinematic equation
$g^{-1}(t)\cdot \dot g(t)=\omega(t)=A(x(t))$.
This problem, for the Fedorov--Kozlov integrable case, 
is studied by Zenkov and Bloch \cite{ZB, ZB1}. 
Similar analyses can be applied to the considered integrable
cases:

{(i)} The periodic solutions on $M$ correspond to
quasi-periodic motions on the phase space $\mathcal M$.

{(ii)} Suppose that in the corollary 3.1 we have 
$c_k=\epsilon C_k$
$k=1,\dots,g$, where $\epsilon$ is a small parameter,
and that the following Diophantine conditions hold:
$$
\vert l+i(k_1\omega_1+\dots+k_g\omega_g)\vert \ge
c(\vert k_1 \vert+\dots+\vert k_g \vert)^{-\gamma}, 
$$
$$
l=0,1,2,
\quad (k_1,\dots,k_g)\in \Bbb{Z}^g -\{0\},
$$
for some constants $c>0$ and $\gamma > g-1$.
Then the reconstruction of quasi-periodic motion (\ref{1.13}) to 
$\mathcal D=\mathcal M$ can
be approximated by quasi-periodic dynamics on the time interval
of length $\sim\exp(1/\epsilon)$.

\section{ Hamiltonian cases}

In some cases, the nonholonomic geodesic flow (\ref{d1.1}) on $\mathcal M$ 
can be gotten as a 
restriction of Hamiltonian flow from $T^*Q$ to 
submanifold $\mathcal M$.
We shall say that nonholonomic geodesic flow (\ref{d1.1}) has a {\it Hamiltonian
restriction property} if there is a function $h^*: T^*Q\to\Bbb{R}$
such that the restriction
of Hamiltonian equation $\dot x=X_{h^*}$ to  $\mathcal M$
coincide with (\ref{d1.1}).

The basic examples are  when Lagrange multipliers in  (\ref{d1.1}) vanish.
Then we can take $h^*(p,q)=h(p,q)=\frac12 p(g_q^{-1}p)$.
However, we shall see that there is another natural choice of the
Hamiltonian $h^*$.
Note that the cases 2 and 4 of example 3.1 have Hamiltonian
restriction property with  Lagrange multipliers 
(in general) different from zero.

From now on, we shall use the notation of section 3.

Let $H(x)=\frac12\langle x,A(x) \rangle$, $x\in G$ 
be the reduced Hamiltonian
of the left-invariant geodesic flow such that
$A=A_L+A_D$ preserve orthogonal decomposition $G=L+D$.
Then  $M=I(D)=D$ and Euler--Poincar\' e--Suslov equations  become:
\begin{equation}
\dot \xi=[\xi,A_D(\xi)]_D, \quad \xi\in D.
\label{6.2}
\end{equation}

Suppose that $L$ is a Lie algebra of some connected Lie subgroup $\mathfrak L$.
Then $[L,L]\subset L$, $[L,D]\subset D$.
Further,
suppose that 
$H_D(\xi)=\frac12\langle \xi,A_D(\xi)\rangle$ is an invariant
of adjoint action of $\mathfrak L$ on $D$.
Recall that function $F:D\to \Bbb{R}$ is $Ad_\mathfrak L$ invariant if 
it satisfy equation:
\begin{equation}
[\xi,\nabla F(\xi)]_L=0, \quad \xi\in D.
\label{6.3}
\end{equation}

Let $x=\xi+\eta$ be orthogonal decomposition of $x\in G$,
such that $\xi\in D$ and $\eta\in L$.
Let $A^*(x)=A^*(\xi+\eta)=A_D(\xi)$.

Define the function $H^*: G \to \Bbb{R}$ by:
\begin{equation}
H^*(x)=\frac12\langle x,A^*(x)\rangle
\label{6.4}
\end{equation}

Using $Ad_\mathfrak L$ invariance of $H_D(\xi)$, 
we are getting that
the Euler--Poincare equations for functions 
$H(x)$ and $H^*(x)$ leave the plain $D$ invariant and on $D$
coincide with  Euler--Poincar\' e--Suslov equations (\ref{6.2}).
For example, the system
\begin{equation}
\dot x=[x,\nabla H^*(x)]=[x,A^*(x)]
\label{6.5}
\end{equation}
after projection to the $L$ and $D$ becomes:
\begin{equation}
\dot \xi=[\xi,A_D(\xi)]+[\eta,A_D(\xi)],
\label{6.6}
\end{equation}
$$
\dot \eta=0.
$$

Let $h,\; h^*:T\mathfrak G\to\Bbb{R}$ be the functions 
obtained by left translations from $H$ and $H^*$.
The nonholonomic geodesic flow has a Hamiltonian
restriction property with functions $h$ and $h^*$.

The Hamiltonian function $h^*$ has a nice geometrical meaning.
Suppose that $D$  generate Lie algebra $G$
by commutations.
Then the distribution $\mathcal D$ is {\it completely nonholonomic}
or {\it bracket generating}.
The Hamiltonian flow $X_{h^*}$ on $T\mathfrak G$ 
is {\it normal sub-Riemannian geodesic
flow} of the sub-Riemannian metric
induced by restriction of the given left-invariant metric to
$\mathcal D$ (for more details see  \cite{St, Ta}).

We can summarize previous considerations in the following theorem:

\begin{theorem}
Suppose that constrained reduced Hamiltonian function
$H_D(\xi)=\frac12\langle \xi,A_D(\xi)\rangle$ is an invariant
of adjoint action of the Lie subgroup $\mathfrak L$ on $D$
and that $D$  generate Lie algebra $G$
by commutations.
Then on the constrained submanifold $\mathcal D$
the following three different problems have the same flow:
nonholonomic geodesic flow, geodesic flow with
Hamiltonian $h$ and sub-Riemannian geodesic flow
with Hamiltonian $h^*$.
\end{theorem}

Recall that equations (\ref{6.5}) are {\it completely integrable}
if they posses a complete involutive  set of integrals.
$\mathcal F=\{F_1,\dots,F_k\}$ is 
{\it complete involutive} set of functions on $G$ if 
$k=\frac12(\dim G+\mathrm{rank}\;G)$,
$dF_1 \wedge \dots \wedge dF_k \ne 0$ on an open dense set 
$\mathcal U$ of $G$ and
functions in $\mathcal F$ are in involution 
with respect to the Lie-Poisson brackets:
$$
\{F_i,F_j\}(x)=\langle x, [\nabla F_j(x),\nabla F_i(x)] \rangle=0, 
\quad {\rm for\; all}\; x\in G.
$$
If the equations  (\ref{6.5}) are completely integrable then
the set $\mathcal U$ is foliated on invariant $(\dim G-k)$--dimensional
tori  with quasi-periodic dynamic
(see \cite{AKN, TF}).
Also,  the system $\dot x=X_{h^*}$ is non-commutatively integrable:
$T\mathfrak G$ is almost everywhere foliated
on invariant isotropic tori (see \cite{MF}).

\begin{remark} {\rm
Integrability of (\ref{6.5})  do not 
implies integrability of the Euler--Poincar\' e--Suslov equations (\ref{6.2}).
As an example, it could be happened that $\mathcal U \cap D =\emptyset$.
If $\mathcal U \cap D$ is an open dense set of $D$, then
the Euler--Poincar\' e--Suslov equations (\ref{6.2}) are also integrable.
In this case, the reconstruction of the flow on the whole phase
space $\mathcal D=\mathcal M$ follows from the reconstruction of 
the flow of Euler--Poincar\' e equations:
the phase space $\mathcal D$ is foliated on invariant tori with
quasi-periodic dynamic.
}
\end{remark}

\begin{example} {\rm
Suppose we are given a chain of connected subgroups:
$$
\mathfrak L=\mathfrak G_0\subset \mathfrak G_1 \subset \dots
\subset  \mathfrak G_n=\mathfrak G
$$
and the corresponding chain of subalgebras:
\begin{equation}
L=G_0\subset G_1 \subset \dots \subset G_n=G.
\label{6.7}
\end{equation}
Let $G_{i}=G_{i-1}+V_i$ be the orthogonal decompositions.
Then
$
D=V_{1}+\dots+V_n.
$
If we define the left-invariant metric by  $A$ of the form:
\begin{equation}
A=A_L+A_D, \quad A_D=s_{1}\cdot Id_{V_{1}}+\dots+s_n\cdot Id_{V_n}, \quad
s_i>0, \; i=1,\dots,n,
\label{6.7a}
\end{equation}
then the function $H_D(\xi)$ will be  an invariant
of adjoint action of $\mathfrak L$ on $D$.
So nonholonomic geodesic flow has a Hamiltonian restriction property.

Suppose that either $(G_i,G_{i-1})$ is a symmetric pair or
$V_i$ is a subalgebra of $G$ for all $i=1,\dots,n$.
Let $\mathcal F_0$ be arbitrary complete commutative
set of functions on the Lie algebra $L$ 
(see constructions in \cite{TF}) lifted to the functions on $G$.
Similarly, if $V_i$ is a Lie subalgebra, 
let  $\mathcal F_i$ be a complete commutative
set of functions on $V_i$ lifted to the functions on $G$.
Otherwise (i.e., if $(G_i,G_{i-1})$ is a symmetric pair) we take
$\mathcal F_i$ to be given by:
$$
\mathcal F_i=\{ f: G \to \mathbb{R}, \; f(x)=p(x_{G_{i-1}}+\lambda x_{V_i}),
\; \lambda\in\mathbb{R},\; p\in\mathcal I(G_i)\},
$$
where $\mathcal{I}(G_i)$ is the
algebra of $Ad_{{\mathfrak G}_i}$ invariant polynomials on $G_i$.

Then it follows from Mikityuk's results (see \cite{Mik}) that
Euler--Poincar\' e equations (\ref{6.5}) 
will be completely integrable.
The complete commutative set of integrals is 
$\mathcal F=\mathcal F_0+\mathcal F_1+\dots+\mathcal F_n$.

For example, let us consider the chains:
$$ 
L= so(k) \subset  so(k+1) \subset \dots \subset so(n),
$$
$$
L=u(k) \subset u(k)+u(1) \subset \dots 
\subset u(n-1)\subset u(n-1)+u(1)\subset u(n),
$$
$$
L=sp(k) \subset sp(k)+sp(1) \subset \dots 
\subset sp(n-1)\subset sp(n-1)+sp(1)\subset sp(n).
$$

Corresponding distributions are completely nonholonomic.
Whence we get integrability of {\it sub-Riemannian geodesic flows}
on Lie groups $SO(n)$, $U(n)$, $Sp(n)$  with left-invariant sub-Riemannian
metrics defined above.

}
\end{example}

\begin{example}{\rm 
Let the Lie subgroup $\mathfrak L$ be commutative.
Then the Lie algebra $L$ is contained in some 
maximal commutative subalgebra $K\subset G$.
Let $D$, $U$ be the orthogonal complements of $L$
in $G$ and $K$ respectively.
Furter, let $D$ generate $G$ by commutations.

Let $a$ be a regular element of $K$.
Then we have 
that $K=\{\eta\in G,\; [\eta,a]=0\}$.
Let $b$ belongs to $K$ and let 
$R: K\to K$ be  symmetric operator
which preserve decomposition $K=L+U$.
By $\varphi_{a,b,R}$ denote operator (so called {\it sectional operator} \cite{TF})
defined with respect to the orthogonal decomposition $G=K+[a,K]$:
$$
\varphi_{a,b,R}\vert_{K}=R,
\quad \varphi_{a,b,R}\vert_{[a,K]}=ad_a^{-1}ad_b.
$$

For compact groups, among sectional operators there are positive
definite. Take such $\varphi_{a,b,R}$.
Let $H(x)= \frac12\langle x,\varphi_{a,b,R}(x)\rangle$.
It can be proved that  
$H_D(\xi)=\frac12\langle \xi,\varphi_{a,b,R}\vert_D(\xi)\rangle$
is an invariant of adjoint action of the Lie subgroup $\mathfrak L$ on $D$.
Thus, nonholonomic geodesic flow has a Hamiltonian restriction property.

The function $H^*$ is of the form:
$H^*(x)= \frac12\langle x,\varphi_{a,b,R^*}(x)\rangle$,
where  $R^*: K\to K$ has a kernel equal to $L$.
For general $a\in K$,
the Euler--Poincar\' e equations (\ref{6.5}) are completely
integrable both on $G$ and on $D$.
The integrals can be obtained by Mishchenko--Fomenko  method,
based on the
shifting of argument of  invariant polynomials
(for details see  \cite{TF}).
Thus corresponding  {\it nonholonomic} and
{\it sub-Riemannian} geodesic flows are integrable.
}
\end{example}

{\bf Acknowledgments.}
The research was carried out while the author was on postdoctoral 
position at the
Graduiertenkolleg "Mathematik in Bereich ihrer Wechselwirkung mit der Physik",
of the Mathematisches Institut, LMU, M\" un\-chen. 
It is a great pleasure to thank
Ludwig Maximilians Universit\"at  for hospitality and
the referees for various useful comments.

\end{document}